\documentclass[
    twocolumn,
	prl,
	amssymb,
	preprintnumbers,superscriptaddress,
	nofootinbib]{revtex4-1}

\pdfoutput=1
\usepackage{paralist}
\usepackage{graphicx}
\usepackage{enumitem}
\usepackage{latexsym}
\usepackage{amsfonts}
\usepackage{amssymb}
\usepackage{amsmath}
\usepackage{xcolor}

\usepackage[normalem]{ulem}
\usepackage{ulem}
\usepackage[
pdfauthor={Djuna Croon}]{hyperref}

\usepackage[export]{adjustbox}
\usepackage[thinlines]{easytable}
\usepackage{slashed}
\usepackage{dcolumn}
\usepackage{verbatim}
\usepackage{float}
\usepackage{multirow}
\usepackage{xspace}
\usepackage{tabularx}
\usepackage{lettrine}
\usepackage{setspace}

\newcommand{\beq}{\begin{equation}}
\newcommand{\eeq}{\end{equation}}
\newcommand{\bea}{\begin{eqnarray}}
\newcommand{\eea}{\end{eqnarray}}

\newcommand{\refjnl}[1]{{\rm#1}}

\def\apj{\refjnl{ApJ}}                 
\def\aap{\refjnl{A\&A}}                

\def\mnras{\refjnl{MNRAS}}             
\usepackage{pdfbase}[2017/03/16]
\usepackage{xparse,ocgbase}
\usepackage{xcolor,calc}
\usepackage{tikzpagenodes,linegoal}
\usetikzlibrary{calc}
\usepackage{tcolorbox}

\ExplSyntaxOn
\let\tpPdfLink\pbs_pdflink:nn
\let\tpPdfAnnot\pbs_pdfannot:nnnn\let\tpPdfLastAnn\pbs_pdflastann:
\let\tpAppendToFields\pbs_appendtofields:n
\def\tpPdfXform{\pbs_pdfxform:nnnnn{1}{1}{}{}}
\let\tpPdfLastXform\pbs_pdflastxform:
\let\cListSet\clist_set:Nn\let\cListItem\clist_item:Nn
\ExplSyntaxOff

\usepackage{pdfbase}[2017/03/16]
\usepackage{xparse,ocgbase}
\usepackage{xcolor,calc}
\usepackage{tikzpagenodes,linegoal}
\usetikzlibrary{calc}
\usepackage{tcolorbox}

\ExplSyntaxOn
\let\tpPdfLink\pbs_pdflink:nn
\let\tpPdfAnnot\pbs_pdfannot:nnnn\let\tpPdfLastAnn\pbs_pdflastann:
\let\tpAppendToFields\pbs_appendtofields:n
\def\tpPdfXform{\pbs_pdfxform:nnnnn{1}{1}{}{}}
\let\tpPdfLastXform\pbs_pdflastxform:
\let\cListSet\clist_set:Nn\let\cListItem\clist_item:Nn
\ExplSyntaxOff

\makeatletter
\NewDocumentCommand{\tooltip}{%
  ssssO{\ifdefined\@linkcolor\@linkcolor\else blue\fi}mO{yellow!20}mO{0pt,0pt}%
}{{%
  \leavevmode%
  \IfBooleanT{#2}{%
    \ocgbase@new@ocg{tipOCG.\thetcnt}{%
      /Print<</PrintState/OFF>>/Export<</ExportState/OFF>>%
    }{false}%
    \xdef\tpTipOcg{\ocgbase@last@ocg}%
    \ocgbase@add@ocg@to@radiobtn@grp{tool@tips}{\ocgbase@last@ocg}%
  }%
  \tpPdfLink{%
    \IfBooleanTF{#4}{%
      /Subtype/Link/Border[0 0 0]/A <</S/SetOCGState/State [/Toggle \tpTipOcg]>>
    }{%
      /Subtype/Screen%
      /AA<<%
        \IfBooleanTF{#3}{%
          /E<</S/SetOCGState/State [/Toggle \tpTipOcg]>>%
        }{%
          \IfBooleanTF{#2}{%
            /E<</S/SetOCGState/State [/ON \tpTipOcg]>>%
            /X<</S/SetOCGState/State [/OFF \tpTipOcg]>>%
          }{
            \IfBooleanTF{#1}{%
              /E<</S/JavaScript/JS(%
                var fd=this.getField('tip.\thetcnt');%
                if(typeof(click\thetcnt)=='undefined'){%
                  var click\thetcnt=false;%
                  var fdor\thetcnt=fd.rect;var dragging\thetcnt=false;%
                }%
                if(fd.display==display.hidden){%
                  fd.delay=true;fd.display=display.visible;fd.delay=false;%
                }else{%
                  if(!click\thetcnt&&!dragging\thetcnt){fd.display=display.hidden;}%
                  if(!dragging\thetcnt){click\thetcnt=false;}%
                }%
                this.dirty=false;%
              )>>%
            }{%
              /E<</S/JavaScript/JS(%
                var fd=this.getField('tip.\thetcnt');%
                if(typeof(click\thetcnt)=='undefined'){%
                  var click\thetcnt=false;%
                  var fdor\thetcnt=fd.rect;var dragging\thetcnt=false;%
                }%
                if(fd.display==display.hidden){%
                  fd.delay=true;fd.display=display.visible;fd.delay=false;%
                }%
               this.dirty=false;%
              )>>%
              /X<</S/JavaScript/JS(%
                if(!click\thetcnt&&!dragging\thetcnt){fd.display=display.hidden;}%
                if(!dragging\thetcnt){click\thetcnt=false;}%
                this.dirty=false;%
              )>>%
            }%
            /U<</S/JavaScript/JS(click\thetcnt=true;this.dirty=false;)>>%
            /PC<</S/JavaScript/JS (%
              var fd=this.getField('tip.\thetcnt');%
              try{fd.rect=fdor\thetcnt;}catch(e){}%
              fd.display=display.hidden;this.dirty=false;%
            )>>%
            /PO<</S/JavaScript/JS(this.dirty=false;)>>%
          }%
        }%
      >>%
    }%
  }{{\color{#5}#6}}%
  \sbox\tiptext{%
    \IfBooleanT{#2}{%
      \ocgbase@oc@bdc{\tpTipOcg}\ocgbase@open@stack@push{\tpTipOcg}}%
    \tcbox[colframe=black,colback=#7,size=fbox,arc=1ex,sharp corners=southwest]{#8}%
    \IfBooleanT{#2}{\ocgbase@oc@emc\ocgbase@open@stack@pop\tpNull}%
  }%
  \cListSet\tpOffsets{#9}%
  \edef\twd{\the\wd\tiptext}%
  \edef\tht{\the\ht\tiptext}%
  \edef\tdp{\the\dp\tiptext}%
  \tipshift=0pt%
  \IfBooleanTF{#2}{%
    \setlength\whatsleft{\linegoal}%
  }{%
    \measureremainder{\whatsleft}%
  }%
  \ifdim\whatsleft<\dimexpr\twd+\cListItem\tpOffsets{1}\relax%
    \setlength\tipshift{\whatsleft-\twd-\cListItem\tpOffsets{1}}\fi%
  \IfBooleanF{#2}{\tpPdfXform{\tiptext}}%
  \raisebox{\heightof{#6}+\tdp+\cListItem\tpOffsets{2}}[0pt][0pt]{%
    \makebox[0pt][l]{\hspace{\dimexpr\tipshift+\cListItem\tpOffsets{1}\relax}%
    \IfBooleanTF{#2}{\usebox{\tiptext}}{%
      \tpPdfAnnot{\twd}{\tht}{\tdp}{%
        /Subtype/Widget/FT/Btn/T (tip.\thetcnt)%
        /AP<</N \tpPdfLastXform>>%
        /MK<</TP 1/I \tpPdfLastXform/IF<</S/A/FB true/A [0.0 0.0]>>>>%
        /Ff 65536/F 3%
        /AA <<%
          /U <<%
            /S/JavaScript/JS(%
              var fd=event.target;%
              var mX=this.mouseX;var mY=this.mouseY;%
              var drag=function(){%
                var nX=this.mouseX;var nY=this.mouseY;%
                var dX=nX-mX;var dY=nY-mY;%
                var fdr=fd.rect;%
                fdr[0]+=dX;fdr[1]+=dY;fdr[2]+=dX;fdr[3]+=dY;%
                fd.rect=fdr;mX=nX;mY=nY;%
              };%
              if(!dragging\thetcnt){%
                dragging\thetcnt=true;Int=app.setInterval("drag()",1);%
              }%
              else{app.clearInterval(Int);dragging\thetcnt=false;}%
              this.dirty=false;%
            )%
          >>%
        >>%
      }%
      \tpAppendToFields{\tpPdfLastAnn}%
    }%
  }}%
  \stepcounter{tcnt}%
}}
\makeatother
\newsavebox\tiptext\newcounter{tcnt}
\newlength{\whatsleft}\newlength{\tipshift}
\newcommand{\measureremainder}[1]{%
  \begin{tikzpicture}[overlay,remember picture]
    \path let \p0 = (0,0), \p1 = (current page.east) in
      [/utils/exec={\pgfmathsetlength#1{\x1-\x0}\global#1=#1}];
  \end{tikzpicture}%
}

\newcommand{\der}{\mathrm{d}}

\newcommand{\msun}{{\rm M}_\odot}

\newcommand{\mbh}{M_{\rm BH}}

\DeclareRobustCommand{\okina}{%
  \raisebox{\dimexpr\fontcharht\font`A-\height}{%
    \scalebox{0.8}{`}%
  }%
}

\allowdisplaybreaks

\makeatletter
\def\blfootnote{\xdef\@thefnmark{}\@footnotetext}
\makeatother

\begin{document}

\title{
Prediction of Multiple Features in the Black Hole Mass Function due to Pulsational Pair-Instability Supernovae
}

\author{Djuna Croon} \email{djuna.l.croon@durham.ac.uk}
\affiliation{Institute for Particle Physics Phenomenology, Department of Physics, Durham University, Durham DH1 3LE, U.K.}
\author{Jeremy Sakstein} \email{sakstein@hawaii.edu}
\affiliation{Department of Physics \& Astronomy, University of Hawai\okina i, Watanabe Hall, 2505 Correa Road, Honolulu, HI, 96822, USA}

\date{\today}

\begin{abstract}
Using high-resolution simulations of black hole formation from the direct collapse of massive stars undergoing pulsational pair-instability supernovae (PPISN), we find a new phenomenon which significantly affects the explosion and leads to two peaks in the resulting black hole mass function (BHMF).~Lighter stars experiencing the pair-instability can form a narrow shell in which alpha ladder reactions take place, exacerbating the effect of the PPISN.~The shell temperature in higher mass stars ($>62 \msun $ at the onset of helium burning for population-III stars with metallicity $Z=10^{-5}$) is too low for this to occur.~As a result, the spectrum of black holes $\mbh (M_i)$ exhibits a shoulder feature whereby a large range of initial masses result in near-identical black hole masses.~PPISN therefore predict two peaks in the mass function of astrophysical black holes --- one corresponding to the location of the upper black hole mass gap and a second corresponding to the location of the shoulder.~This shoulder effect may explain the peak at $35_{-2.9}^{+1.7}{\rm M}_\odot$ in the LIGO/Virgo/KAGRA GWTC-3 catalog of merging binary black holes.
\end{abstract}

\preprint{IPPP/23/62}

\maketitle

The\blfootnote{\textbf{Author Contribution Statement:}~Both authors contributed equally to all aspects of this manuscript.~The author list is alphabetical.} 
observation of merging binary black holes (BBHs) in gravitational waves by LIGO/Virgo/KAGRA has enabled us to probe their population statistics.~A large number of these objects are expected to have formed by the direct collapse of massive stars ($M\ge20\msun$).~The shape of the black hole mass function (BHMF) as well as features such as the presence of mass gaps provides insights into the structure and evolution of these massive star progenitors.~For example, the upper black hole mass gap, which has been detected in the data\cite{Fishbach:2017zga,LIGOScientific:2018jsj,Baxter:2021swn}, is the result of pair-instability supernovae\cite{1967ApJ...148..803R,Woosley:2016hmi}, and its location provides information about nuclear reaction rates, metallicity, and mass-loss\cite{Farmer:2019jed,Farmer:2020xne,Baxter:2021swn,Mehta:2021fgz}.

The BHMF inferred from the most recent GWTC-3 catalog exhibits a peak at $35_{-2.9}^{+1.7}\msun$ (error bars indicate the 90\% confidence interval of the center of the Gaussian peak in the powerlaw+peak model) for which no physical origin is known\cite{LIGOScientific:2021psn}.~This peak is detected with high significance using both parametric and non-parametric data analyses\cite{LIGOScientific:2021psn,Edelman:2021zkw,Edelman:2022ydv,Callister:2023tgi,Farah:2023swu}.~Finding the mechanism responsible for the formation of this peak would provide new insights into the relevant astrophysics, broadening the science accessible to terrestrial gravitational wave interferometers.~

The only known mechanism to explain a high-mass feature in the BHMF is  pulsational pair-instability supernovae (PPISN).~Massive stars may encounter the \textit{pair-instability} where their interior temperatures and densities are sufficient to thermally produce electron-positron pairs from energetic photons.~This results in a contraction that is ultimately halted by explosive oxygen ignition.~The explosion drives a series of mass-shedding pulsations.~As a result, 
a broad range of initial stellar masses ultimately collapse to form similar mass black holes, leading to a sharp peak in the BHMF referred to as \textit{the upper black hole mass gap}.~The PPISN peak has been investigated in many previous works\cite{Woosley:2016hmi,Marchant:2020haw,Farmer:2020xne,Mehta:2021fgz}, all of which predict it to lie at  higher masses ($\mbh \sim 46-93 \, \msun$) than the observation in the GWTC-3 catalog\cite{LIGOScientific:2021psn}.~This is true even when nuclear, astrophysical, and modelling uncertainties are taken into account\cite{Farmer:2019jed,Farmer:2020xne,Mehta:2021fgz}.~

In this work, we present the results of high-resolution simulations of the stage immediately preceding black hole formation from the collapse of massive stars experiencing PPISN.~Through this analysis, we find evidence of a shoulder in $\mbh( M_{i}) $ where $M_i$ is the mass of the star at the onset of helium burning --- where we begin our simulations --- implying that the BHMF exhibits a second peak.~The astrophysics responsible for the second peak is the formation of a narrow convective shell, in which carbon burning enables the $\alpha$-ladder reactions, predominantly 
$\rm ^{16}O (\alpha,\gamma)^{20}Ne $, to generate large energies.~We find that this effect exacerbates the PPISN explosion for low-mass stars, but is not present for higher-mass stars.
High resolution simulations are required to resolve the convective shell burning responsible for this feature.

\begin{figure*}
    \centering
\includegraphics[width=.6\textwidth]{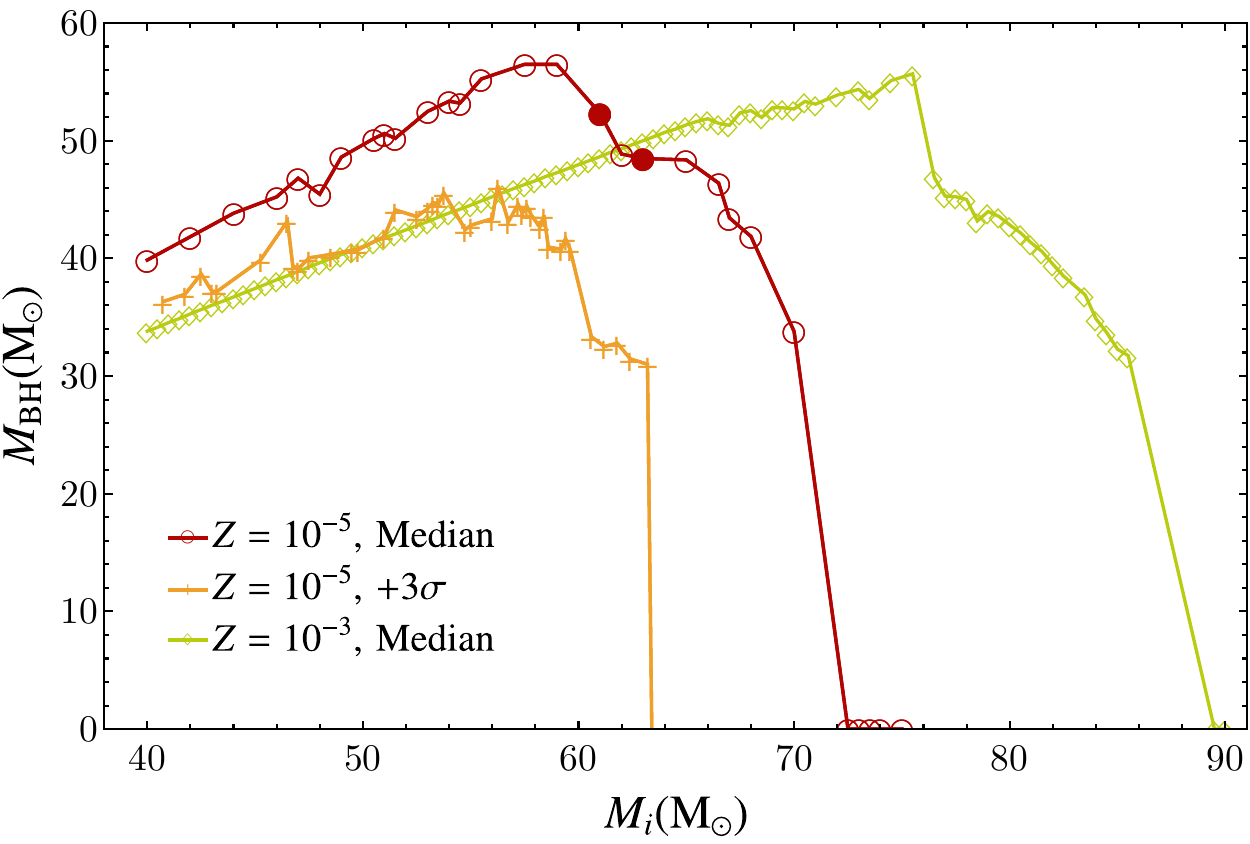}
    \caption{Grids of results for different values of the metallicity $Z$ and the $\rm ^{12}C(\alpha,\gamma)^{16}O $ rate.~The filled dots indicate the two stars (one preceding the shoulder and one in the shoulder) used for comparison in later plots.}
    \label{fig:shoulder}
\end{figure*}

For low metallicity $Z= 10^{-5}$ (population-III) stars and median $^{12}{\rm C}(\alpha,\gamma)^{16}{\rm O}$ rate\cite{deBoer:2017ldl}, we find that the new peak appears at $\mbh \sim 48 \msun$.~However, this peak may be shifted as a function of other uncertainties.~To demonstrate this, we show that a $3\sigma$ increase in the $^{12}{\rm C}(\alpha,\gamma)^{16}{\rm O}$ rate brings the peak to $\sim 33\msun$.~We therefore propose that the exacerbation of  PPISN by convective carbon burning shells in lower mass stars, absent in higher mass stars, is a good candidate for the origin of the $35\msun$ peak observed in the LIGO/Virgo/KAGRA black hole mass function.

Our simulations were preformed using the one-dimensional stellar structure code MESA version 12778.~The full details of how PPISN are treated by MESA can be found in\cite{Paxton:2017eie,Marchant:2018kun,Farmer:2019jed}.~Here we recap only the salient features.~MESA is equipped with a hydrodynamic solver that enables it to go beyond hydrostatic equilibrium and track the evolution of the star through the PPISN phase\cite{1994ShWav...4...25T}.~The solver can model shocks and pulsations while conserving energy to high accuracy.~After the pulsations end, the bound layers of the star return to hydrostatic equilibrium while the ejected layers expand and cool.~MESA cannot simultaneously track the evolution of the bound star and the ejected layers because the latter leave the range of applicability of the MESA equation of state.~To avoid issues associated with this, the outer layers are removed using a relaxation process that creates a hydrostatic model that matches the mass, composition, and entropy of the bound layers (see\cite{Paxton:2017eie} Appendix B and\cite{Marchant:2018kun} Appendix C for more details).~We begin our simulations with helium cores on the zero-age helium-branch (ZAHB) since the hydrogen envelope is expected to be removed by binary interactions\cite{Kobulnicky:2006bk,Sana:2012px,2017A&A...598A..84A}, strong stellar winds\cite{Vink:2005zf,2017A&A...603A.118R}, opacity-driven mass loss\cite{2015A&A...573A..18M}, or chemically homogeneous binary evolution leading to rapid rotation\cite{Maeder:2000wv,Yoon:2006fr,deMink:2009jq,Mandel:2015qlu,Marchant:2016wow}.~
We include mass loss according to the prescription of\cite{Brott:2011ni}, which is appropriate for these objects.~We use the default MESA nuclear reaction rates with the exception of the $^{12}{\rm C}(\alpha,\gamma)^{16}{\rm O}$ rate, which is the most important source of uncertainty effecting the PPISN\cite{Farmer:2019jed,Farmer:2020xne,Mehta:2021fgz}.~We use the state-of-the-art rate calculated by\cite{deBoer:2017ldl}.~Compared with previous works\cite{Mehta:2021fgz}, our simulations increase the number of cells by a factor of $\sim1.6$ ({\sc mesh\_delta\_coeff = 0.5}) and reduces the correction scale factor for temperature corrections by a factor of 20 ({\sc scale\_max\_correction = 0.05}).~The complete set of parameters we adopt can be found in the reproduction package that will be made public upon publication of this paper.~

As discussed at length in\cite{Farmer:2019jed,Farmer:2020xne}, the PPISN phase is highly sensitive to the $^{12}{\rm C}(\alpha,\gamma)^{16}{\rm O}$ rate.~Decreasing this rate
causes less carbon to be processed into oxygen during core helium burning, resulting in the formation of a convective carbon burning shell before the pair-instability.~This shell resists the contraction and therefore dampens the PPISN.~Conversely, increasing the rate results in more carbon being converted to oxygen, preventing the formation of the convective shell  and ultimately intensifying the explosion.~The $^{12}{\rm C}(\alpha,\gamma)^{16}{\rm O}$ rate  is highly uncertain due to its theoretical and experimental inaccessibility\cite{deBoer:2017ldl}.~Variations in this rate of up to $3 \sigma$ from the median have been studied in a range of stellar objects\cite{2016MNRAS.456.3866C,deBoer:2017ldl,Farmer:2019jed,Farmer:2020xne}.~In what follows, we will vary this rate to determine its effect on the location of the second peak.~Varying other parameters e.g., the mass loss efficiency and mixing length will move the location of the peak in a sub-dominant and predictable manner.

\begin{figure*}
    \centering
    \includegraphics[height=6.1cm]{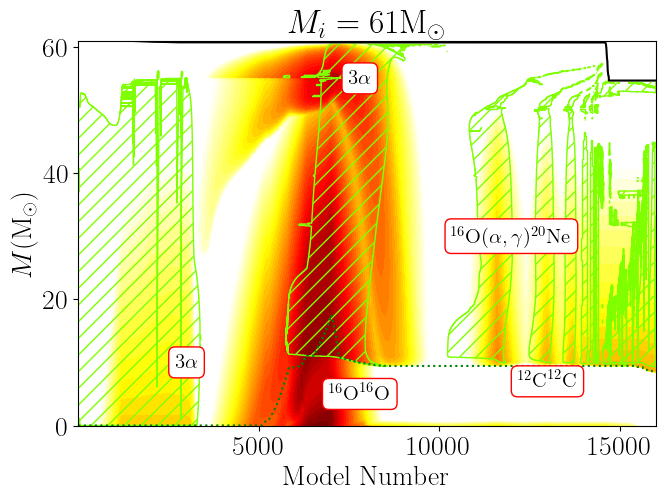}
    \includegraphics[height=6.1cm]{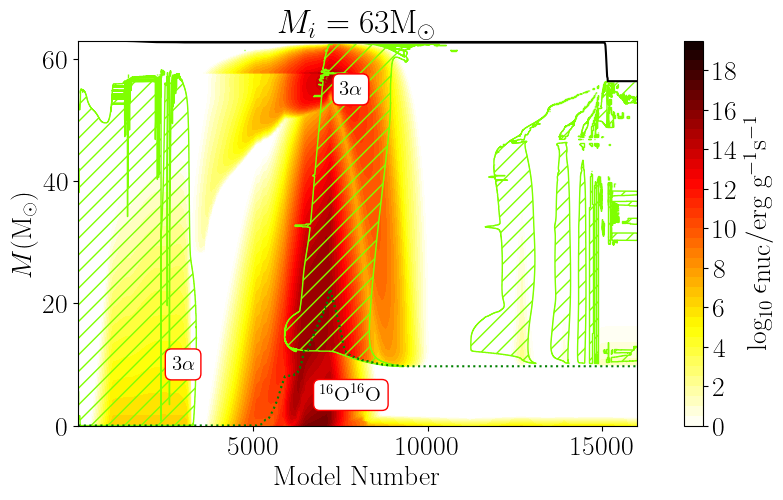}
    \caption{Kippenhahn diagram of the nuclear burning rate of a pre-shoulder star ($61\msun$, left panel) and a star in the shoulder ($63 \msun$, right panel) for $Z=10^{-5}$ and the median $^{12}C(\alpha,\gamma)^{16}O $ rate.~Model number non-linearly corresponds to time in the simulation.~Convective zones are indicated by the dashed green regions.~The dominant rates are labeled in the plot.~The shell-burning episodes starting at model number $\sim11,500$ are absent in shoulder stars.}
    \label{fig:kips}
\end{figure*}

The results of our simulations are shown in figure \ref{fig:shoulder} where we plot the mass of the black hole formed as a function of the initial ZAHB mass for both the median $^{12}{\rm C}(\alpha,\gamma)^{16}{\rm O}$ rate and the $+3\sigma$ deviation from this for our default model with $Z=10^{-5}$ as can be expected for population-III stars.~We also show results for higher metallicity $Z=10^{-3}$ objects.~The upper black hole mass gap is evident by the lack of black holes with masses heavier than $56.5\msun$ for the median $^{12}{\rm C}(\alpha,\gamma)^{16}{\rm O}$ rate and $46.2\msun$ for the $^{12}{\rm C}(\alpha,\gamma)^{16}{\rm O}$$+3\sigma$ rate (values correspond to the $Z=10^{-5}$ predictions).~This feature is well understood at the result of PPISN.~Also evident is a shoulder feature where stars with a broad range of initial masses form similar mass black holes:~$48\msun$ in the case of the median rate and $33\msun$ in the case of the $+3\sigma$ rate.~One therefore expects a peak in the BHMF at the shoulder mass --- an expectation that we will confirm quantitatively below.

\begin{figure*}
    \centering
    \includegraphics[height=6.1cm]{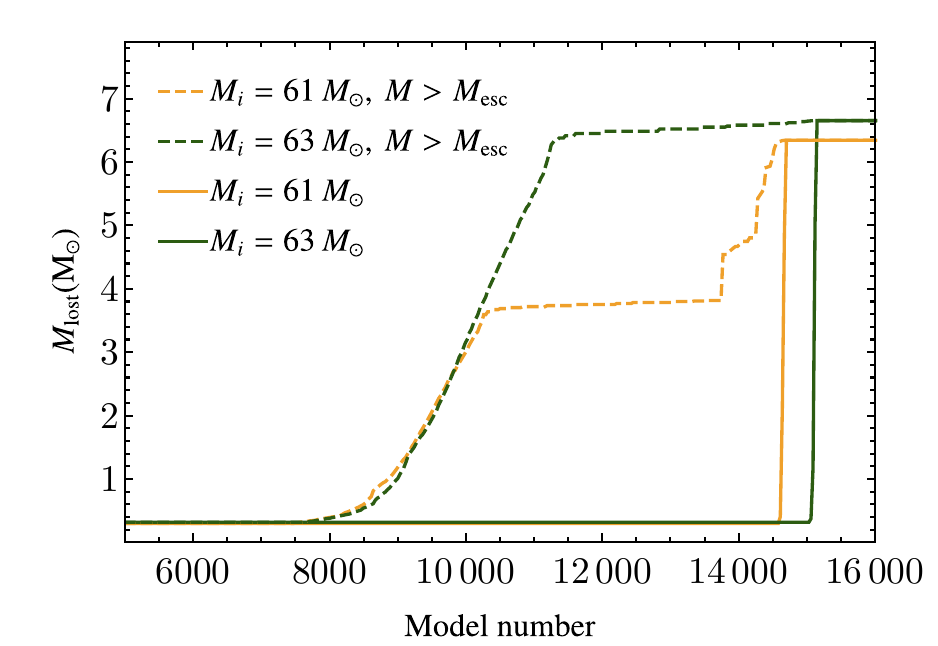}
    \includegraphics[height=6.1cm]{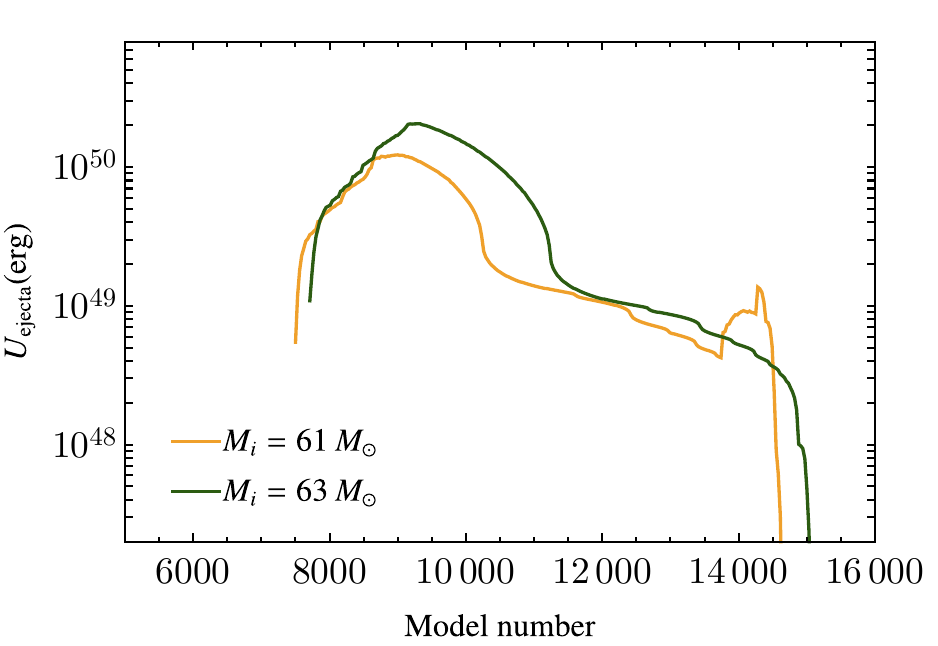}
    \caption{\textit{Left:} Mass lost from a pre-shoulder ($61\msun$) and shoulder ($63\msun$) star during the PPISN phase as a function of model number.~Dashed lines show the unbound mass (mass of layers expanding faster than escape velocity);~solid lines show the mass removed from the star.~The solid and dashed lines coincide at the end of the PPISN.~\textit{Right:}~Internal energy of the ejected material for a pre-shoulder ($61\msun$) and shoulder ($63\msun$) star.~The $\alpha$-ladder shell burning in the pre-shoulder star around model number 13,500 injects energy into the outer layers, causing the second episode of mass loss seen in the left panel.~This injection is absent in the shoulder star. }
    \label{fig:vesc}
\end{figure*}

The physical origin of the shoulder can be discerned by examining the Kippenhahn plots in Fig.~\ref{fig:kips}, which show the nuclear burning rate $\epsilon_{\rm nuc}$ in the star before the first and largest loss of mass.~We compare a star with initial mass $61 \msun$, which is prior to the shoulder, and one with initial mass $63 \msun$, which is part of it.~Both stars are labeled in Fig.~\ref{fig:shoulder}.
In Fig.~\ref{fig:kips}, one can clearly see that the $61\msun$ star undergoes a number of shell burning episodes after the initial phase of explosive core oxygen burning, which are absent in the $63\msun$ star.~The reason for this is the following.~After the explosive burning phase but before relaxation, pre-shoulder stars reach inner shell temperatures of $\sim 6 \times 10^8 \rm K$, at which carbon burning (indicated by $\rm ^{12}C^{12}C$ in Fig.~\ref{fig:kips}) takes place.~The alpha particle created in this reaction can subsequently be captured as part of the alpha ladder:
\begin{equation}
    \begin{split}
        \rm ^{16}O + \alpha &\to \, \rm ^{20}Ne + \gamma \\
        \rm ^{20}Ne + \alpha &\to \, \rm ^{24}Mg + \gamma \\
        \rm ^{24}Mg + \alpha &\to \, \rm ^{28}Si + \gamma 
    \end{split}
\end{equation}
with the first reaction being dominant.~Each alpha capture leads to the release of $4.5-10 \rm MeV$.~
This energy release accelerates the stellar material and importantly, unbinds a large fraction of the previously bound mass in the outer layers of the star, an effect we display in Fig.~\ref{fig:vesc}.~Thus, the pair instability explosion is exacerbated by the alpha process in these stars.~In contrast, in the shoulder stars, the temperatures in the shell do not reach sufficiently high values for the onset of this chain of events.~The reason that the heavier stars do not reach the requisite temperature for carbon burning is that pair-instability is stronger in heavier stars, resulting in more violent initial explosions and, consequentially, more cooling.

\begin{figure*}
    \centering
    \includegraphics[width=\textwidth]{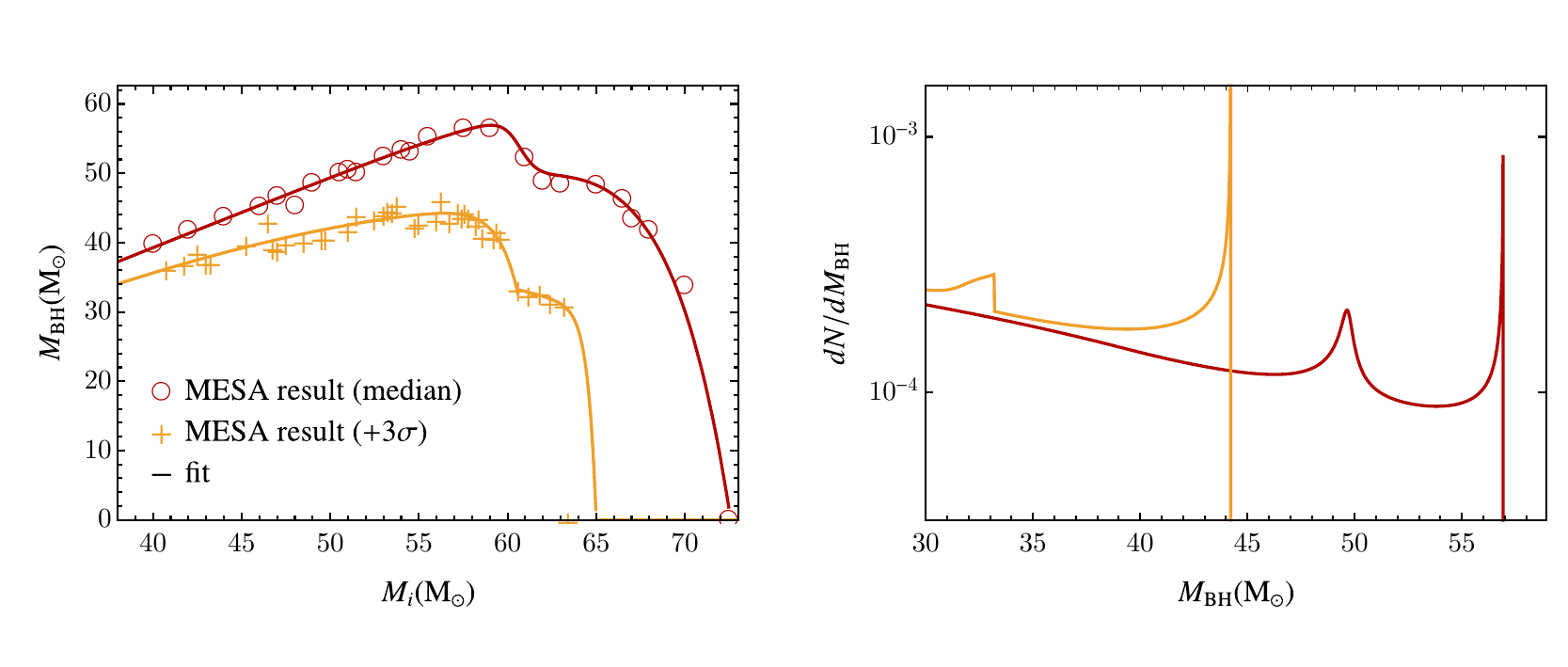}
    \caption{Implications of our results for the black hole mass function observable via gravitational waves, predicted using a fit to our simulation results. The peak in the $ +3\sigma $ has a sharp drop off due to the discontinuity in $\mbh(M_i)$.~Here we assumed an initial mass function $\der N/ \der M_{\rm ZAHB}\propto M_i^{-2.4} $.
    }
    \label{fig:BHMFs}
\end{figure*}

We now derive the effects of the alpha processes on the BHMF, which is observed via gravitational radiation.~As a function of successive stellar epochs BHMF is given by\cite{Baxter:2021swn}
\begin{equation}
    \frac{\der N}{\der M_{\rm BH}}=\frac{\der N}{\der M_{\rm ZAMS}}\frac{\der M_{\rm ZAMS}}{\der M_{\rm ZAHB}}\left(\frac{\der M_{\rm BH}}{\der M_{\rm ZAHB}}\right)^{-1},
    \label{eq:BHMF}
\end{equation}
where ZAMS refers to the zero-age main-sequence.~The first term on the right is the initial mass function (IMF), which is typically assumed to be a power-law with negative slope\cite{1955ApJ...121..161S};~the second term accounts for the effects of mass loss on the main-sequence before the ZAHB and can also reasonably be assumed to be a power-law\cite{Baxter:2021swn};~and the third term is the inverse of the  derivative of the curve(s) shown in figure~\ref{fig:shoulder}.~The derivative ${\der M_{\rm BH}}/{\der M_{\rm ZAHB}}$ is zero at the location of the upper black hole mass gap, leading to a divergence in the BHMF at this mass.~This is expected to manifest as a peak in the BHMF due to noise and a finite sample size.~At the shoulder, ${\der M_{\rm BH}}/{\der M_{\rm ZAHB}}$ levels off to near zero values before increasing again, leading to a peak feature in the BHMF.~We show this in Fig.~\ref{fig:BHMFs} where we fit an interpolating function to the results of our simulation and use this in equation \eqref{eq:BHMF} to derive the BHMF. 

Fig.~\ref{fig:BHMFs} demonstrates that
the $\sim35\msun$ peak can be accommodated by the shoulder feature if we allow the other parameters to deviate from their median values.~Most importantly, if the rate of the $^{12}{\rm C}(\alpha,\gamma)^{16}{\rm O}$ is $3\sigma$ larger than its median value the second peak would be located at $\mbh \sim 33 \msun$.~This predicts the location of the upper mass gap to be $M_{\rm BHMG}=46.2\msun$, implying that all objects in the GWTC-3 catalog heavier than this are either second generation black holes formed from via mergers, or experienced an extended stage of mass accretion e.g., the black hole could have been formed in an AGN disk and accreted the surrounding material (see\cite{LIGOScientific:2020ufj} for various mechanisms for populating the mass gap).~Previous works have investigated whether the $35\msun$ peak can be explained by PPISN, and have concluded that it can not\cite{Hendriks:2023yrw,Golomb:2023vxm}.~These works used parametric formulas for the black hole mass spectrum $M_{\rm BH}(M_{\rm ZAHB})$ that do not account for the shoulder feature that we find, and hence are unable to accommodate a second peak.

Our discovery that PPISN predict multiple features in the BHMF have broad implications for gravitational wave astronomy and cosmology that we discuss below.~Before doing so, we discuss the limitations of our present PPISN simulations.~First, we have not simulated rotating objects.~Rotation both reduces the region in the central temperature--central density plane where the pair-instability occurs and causes the star to evolve away from this region, resulting in more massive black holes formed\cite{Marchant:2020haw}.~The majority of stars are expected to be only slowly-rotating due to the Spruit-Tayler dynamo\cite{1999A&A...349..189S,2002A&A...381..923S}, so we anticipate that rotation will cause a small scatter in the location of the shoulder, which in turn would make the peak noisy.~A second limitation is that we have simulated isolated objects.~The black holes observed by LIGO/Virgo/KAGRA are expected to have formed in binary star systems.~To date, there have not been any simulations of PPISN in binary systems, so it is harder to predict how binarity would effect the shoulder feature.~Finally, MESA is a one-dimensional code.~Three-dimensional processes are important during non-hydrostatic phases such as PPISN.~These may alter the properties of the shoulder.~We believe that investigations of binary and 3D effects on PPISN are warranted in  light of our findings, and their potential implications for stellar and gravitational wave astronomy.

Our simulations have revealed a shoulder feature in the black hole mass spectrum $\mbh(M_i)$ --- the black hole mass as a function of zero age helium burning mass -- where stars with a range of initial masses collapse to form near-identical mass black holes.~The shoulder is due to the absence of post-explosion alpha ladder processes in the convective shells of these objects.~We have demonstrated that this shoulder gives rise to a second peak in the black hole mass function in addition to the expected peak preceding the upper black hole mass gap peak at higher masses.~The $35\msun$ peak in the black hole mass function observed by LIGO/Virgo/KAGRA can be explained if the (highly uncertain) $^{12}{\rm C}(\alpha,\gamma)^{16}{\rm O}$ rate is larger than its median value.

To date, no convincing explanation for the $35\msun$ peak has been found.~We believe that further investigations of our proposal are therefore warranted.~These include investigating the properties/existence of the shoulder beyond our current isolated, non-rotating, one-dimensional simulations;~and testing its prediction that  LIGO/Virgo/KAGRA black holes heavier than $46.2\msun$ were formed via merger/accretion channels.~

Our result indicates that PPISN predicts a doubly peaked spectrum of black holes in the gravitational wave data. The $35\msun$ peak detected with high statistical significance in GWTC-3 is a good candidate for the first peak; its true location can be determined decisively through future detections.~Measuring the location of this second peak will aid us in interpreting gravitational wave catalogs to understand the astrophysical processes governing the universe.~Specifically, the knowledge of the location, height, and width of the peak will help to constrain stellar and nuclear physics such as the $^{12}{\rm C}(\alpha,\gamma)^{16}{\rm O}$ rate and the wind loss efficiency.~In addition, features in the BHMF can break degeneracies between black hole and cosmological parameters, enabling measurements of the Hubble constant and dark matter density\cite{Ezquiaga:2022zkx}.~The second peak we have predicted provides additional information that may help to reduce the (currently large) error bars in these measurements.

\section*{Software}
MESA version~12778, MESASDK version 20200325,  
mkipp\footnote{\href{https://github.com/orlox/mkipp}{https://github.com/orlox/mkipp}}, 
Mathematica version 12.

\section*{Acknowledgements}

We thank Ebraheem Farag and Frank Timmes for helpful discussions.~DC is supported by the STFC under Grant No.~ST/T001011/1.~This material is based upon work supported by the National Science Foundation under Grant No.~2207880.~Our simulations were run on the University of Hawai\okina i's high-performance supercomputer KOA.~The technical support and advanced computing resources from University of Hawai\okina i Information Technology Services – Cyberinfrastructure, funded in part by the National Science Foundation MRI award \#1920304, are gratefully acknowledged.~

\bibliography{refs}

\end{document}